\documentclass[prb,preprint,showpacs,preprintnumbers,amsmath,aps]{revtex4}
\usepackage{graphicx,hyperref}% Include figure files
\usepackage{bm}% bold math
\usepackage{subfigure}% bold math

\def\nbC{{\mathchoice {\setbox0=\hbox{$\displaystyle\rm C$}%
\hbox{\hbox to0pt{\kern0.4\wd0\vrule height0.9\ht0\hss}\box0}}
{\setbox0=\hbox{$\textstyle\rm C$}\hbox{\hbox
to0pt{\kern0.4\wd0\vrule height0.9\ht0\hss}\box0}}
{\setbox0=\hbox{$\scriptstyle\rm C$}\hbox{\hbox
to0pt{\kern0.4\wd0\vrule height0.9\ht0\hss}\box0}}
{\setbox0=\hbox{$\scriptscriptstyle\rm C$}\hbox{\hbox
to0pt{\kern0.4\wd0\vrule height0.9\ht0\hss}\box0}}}}
\def\nbQ{{\mathchoice {\setbox0=\hbox{$\displaystyle\rm
Q$}\hbox{\raise
0.15\ht0\hbox to0pt{\kern0.4\wd0\vrule height0.8\ht0\hss}\box0}}
{\setbox0=\hbox{$\textstyle\rm Q$}\hbox{\raise
0.15\ht0\hbox to0pt{\kern0.4\wd0\vrule height0.8\ht0\hss}\box0}}
{\setbox0=\hbox{$\scriptstyle\rm Q$}\hbox{\raise
0.15\ht0\hbox to0pt{\kern0.4\wd0\vrule height0.7\ht0\hss}\box0}}
{\setbox0=\hbox{$\scriptscriptstyle\rm Q$}\hbox{\raise
0.15\ht0\hbox to0pt{\kern0.4\wd0\vrule height0.7\ht0\hss}\box0}}}}
\def\nbT{{\mathchoice {\setbox0=\hbox{$\displaystyle\rm
T$}\hbox{\hbox to0pt{\kern0.3\wd0\vrule height0.9\ht0\hss}\box0}}
{\setbox0=\hbox{$\textstyle\rm T$}\hbox{\hbox
to0pt{\kern0.3\wd0\vrule height0.9\ht0\hss}\box0}}
{\setbox0=\hbox{$\scriptstyle\rm T$}\hbox{\hbox
to0pt{\kern0.3\wd0\vrule height0.9\ht0\hss}\box0}}
{\setbox0=\hbox{$\scriptscriptstyle\rm T$}\hbox{\hbox
to0pt{\kern0.3\wd0\vrule height0.9\ht0\hss}\box0}}}}
\def\nbS{{\mathchoice
{\setbox0=\hbox{$\displaystyle     \rm S$}\hbox{\raise0.5\ht0%
\hbox to0pt{\kern0.35\wd0\vrule height0.45\ht0\hss}\hbox
to0pt{\kern0.55\wd0\vrule height0.5\ht0\hss}\box0}}
{\setbox0=\hbox{$\textstyle        \rm S$}\hbox{\raise0.5\ht0%
\hbox to0pt{\kern0.35\wd0\vrule height0.45\ht0\hss}\hbox
to0pt{\kern0.55\wd0\vrule height0.5\ht0\hss}\box0}}
{\setbox0=\hbox{$\scriptstyle      \rm S$}\hbox{\raise0.5\ht0%
\hboxto0pt{\kern0.35\wd0\vrule height0.45\ht0\hss}\raise0.05\ht0%
\hbox to0pt{\kern0.5\wd0\vrule height0.45\ht0\hss}\box0}}
{\setbox0=\hbox{$\scriptscriptstyle\rm S$}\hbox{\raise0.5\ht0%
\hboxto0pt{\kern0.4\wd0\vrule height0.45\ht0\hss}\raise0.05\ht0%
\hbox to0pt{\kern0.55\wd0\vrule height0.45\ht0\hss}\box0}}}}
\def\nbZ{{\mathchoice {\hbox{$\sf\textstyle Z\kern-0.4em Z$}}
{\hbox{$\sf\textstyle Z\kern-0.4em Z$}}
{\hbox{$\sf\scriptstyle Z\kern-0.3em Z$}}
{\hbox{$\sf\scriptscriptstyle Z\kern-0.2em Z$}}}}
\begin{document}

\title{Critical behavior of the random-field Ising model with long-range 
interactions in one dimension}

\author{Ivan Balog} \email{balog@ifs.hr}
\affiliation{Institute of Physics, P.O. Box 304, Bijeni\v{c}ka cesta 46, 
HR-10001 
Zagreb, Croatia}

\author{Gilles Tarjus} \email{tarjus@lptmc.jussieu.fr}
\affiliation{LPTMC, CNRS-UMR 7600, Universit\'e Pierre et Marie Curie,
bo\^ite 121, 4 Pl. Jussieu, 75252 Paris c\'edex 05, France}

\author{Matthieu Tissier} \email{tissier@lptmc.jussieu.fr}
\affiliation{LPTMC, CNRS-UMR 7600, Universit\'e Pierre et Marie Curie,
bo\^ite 121, 4 Pl. Jussieu, 75252 Paris c\'edex 05, France}

\date{\today}

\begin{abstract}
We study the critical behavior of the one-dimensional random field Ising model 
(RFIM) with long-range 
interactions ($\propto r^{-(d+\sigma)}$) by the nonperturbative functional 
renormalization group. We find two distinct regimes of critical behavior as a 
function of 
$\sigma$, separated by a critical value $\sigma_c$. What distinguishes these 
two regimes is the presence or 
not of a cusp-like nonanalyticity in the functional dependence of the 
renormalized cumulants of the 
random field at the fixed point. This change of behavior can be associated to 
the characteristics of the 
large-scale avalanches present in the system at zero temperature. We propose 
ways to check these 
predictions through lattice simulations. We also discuss the difference with 
the RFIM on the Dyson hierarchical lattice.
\end{abstract}

\maketitle

\section{Introduction}

The random field Ising model (RFIM) has been the focus of intense investigation 
as one 
of the paradigms of criticality in the presence of quenched 
disorder.\cite{imry-ma75,nattermann98} It 
has found application in physics and physical chemistry as well as in 
interdisciplinary fields such as biophysics, 
socio- and econo-physics. The model displays a phase transition, a 
paramagnetic-to-ferromagnetic  one in the language of 
magnetic systems, and the long-distance physics is dominated by 
disorder-induced sample-to-sample 
fluctuations rather than by thermal fluctuations. In the renormalization-group 
(RG) sense, the critical behavior 
is then controlled by a fixed point at zero temperature and its universal 
properties can also be studied 
by investigating the model at zero temperature as a function of disorder 
strength.

One of the central issues arising in the RFIM was the so-called 
dimensional-reduction property, according 
to which the critical behavior of the random system is the same as that of the 
pure system 
in a dimension reduced by 2. This was found at all orders of perturbation 
theory\cite{grin_76,young_77} and was 
related to the presence of an underlying supersymmetry.\cite{parisi_79}. The 
property fails in low dimension, 
in particular in $d=3$ \cite{imbrie_84,brich_kup_87} and a resolution of the 
problem was found within the framework 
of the nonperturbative functional RG 
(NP-FRG).\cite{tarjus04,tissier06,tissier11} Within the NP-FRG, the breakdown 
of 
dimensional reduction and the associated spontaneous breaking of the underlying 
supersymmetry are attributed 
to the appearance of a strong enough nonanalytic dependence, a ``cusp'', in the 
dimensionless renormalized cumulants 
of the random field at the fixed point.

What appears specific to random-field systems among the disordered models whose 
long-distance behavior is 
controlled by a zero-temperature fixed point for which perturbation theory 
predicts the $d \rightarrow d-2$ 
dimensional-reduction property, as, \textit{e.g.}, interfaces in a random 
environment,\cite{fisherFRG,
FRGledoussal-chauve,FRGledoussal-wiese} is the existence of two 
distinct regimes of the critical behavior separated by a \textit{nontrivial} 
value of the dimension,\cite{tarjus04,tissier11} 
or of the number of components in the O(N) model,\cite{tissier06} or else of 
the power-law exponent of the interactions 
for the long-range models.\cite{bacz_LR3d} This is actually what requires a RG 
treatment that should be not only 
functional but also nonperturbative (hence, the NP-FRG).

It was shown in Ref. [\onlinecite{tarj_avlnch}] that these two regimes are 
related to the large-scale properties of 
the ``avalanches'', which are collective phenomena present at zero temperature. 
In equilibrium, such ``static'' 
avalanches describe the discontinuous change in the ground state of the system 
at values of the external source 
that are sample-dependent. At the critical point, the avalanches always take 
place on all scales. However, whether 
or not they induce a ``cusp'' in the dimensionless renormalized cumulants of 
the random field at the fixed point 
depends on their scaling properties, and more specifically on the fractal 
dimension $d_f$ of the largest typical 
avalanches at criticality compared to the scaling dimension of the total 
magnetization.\cite{tarj_avlnch} 

In the short-range RFIM the change of critical behavior appears in a large, 
noninteger dimension 
$d\approx 5.1$,\cite{tarjus04,tissier06,tissier11} which 
is therefore not accessible to lattice simulations. The interest of introducing 
long-range interactions is  that this  
reduces the dimensions where a phase transition can be observed and provides an 
additional control parameter with 
the power-law exponent governing the spatial decay of the interactions. In Ref. 
[\onlinecite{bacz_LR3d}], 
a $3$-dimensional RFIM with both long-range interactions and long-range 
correlations of the random field was studied. The long-distance decays were 
chosen in such a way 
that the supersymmetry which is responsible for dimensional reduction is still 
present in the theory. It was then shown 
that the spontaneous breaking of supersymmetry and the associated breakdown of 
dimensional reduction 
do take place in this $3-d$ model and that two regimes of critical behavior are 
present and separated by a nontrivial value 
of the power-law exponents.\cite{bacz_LR3d}

In this work, we further reduce the dimension of interest to $d=1$ by 
considering the RFIM with 
long-range interactions but short-range correlations of the disorder. This 
model should be even more 
accessible to lattice simulations (see for instance the recent studies in Refs. 
[\onlinecite{leuzzi,dewenter}]). 
This should allow an independent check of the scenario derived from the NP-FRG. 

In the model the interactions decay with distance as $r^{-(d+\sigma)}$ with 
$\sigma >0$. In some sense, varying 
$\sigma$ has a similar effect to that of changing the spatial dimension in the 
short-range case. For 
$\sigmaÊ\leq \sigma_G=1/3$, the critical behavior is governed by a Gaussian 
fixed point and the 
exponents are therefore the classical (mean-field) ones in the presence of 
long-range interactions: $\sigma_G$ is 
the analog of an upper critical dimension. On the other hand, heuristic 
arguments predict that no 
phase transition takes place for $\sigma \geq \sigma_M=1/2$:\cite{Bray,leuzzi} 
$\sigma_M$ is then 
the analog of an upper critical dimension. The interesting range is therefore 
$1/3\leq \sigma <1/2$. 
(Rigorous results show the existence of a phase transition for 
$2-(\ln3/\ln2)\approx 0.4150\cdots <\sigma 
< 1/2$\cite{cassandro09} and it is even more likely that a transition exists 
for smaller values of $\sigma$.)

We investigate through the NP-FRG whether this $1$-dimensional long-range RFIM 
displays, as the other 
random-field models studied so far, two regimes of critical behavior separated 
by a nontrivial value, here of 
$\sigma$. In this model, there is of course no $d \rightarrow d-2$ 
dimensional-reduction property and no 
associated supersymmetry. The two regimes should therefore be characterized by 
the presence or not 
of a ``cusp'' in the functional dependence of the (dimensionless) renormalized 
cumulants of the random field 
at the fixed point.

We do find two distinct regimes separated by the critical value 
$\sigma_c\approx 0.379$: a  ``cuspless'' one 
below $\sigma_c$ and a ``cuspy'' one above. We calculate the critical 
exponents, which do not show any 
significant change of behavior around $\sigma_c$. More significant is the 
variation of the fractal 
dimension $d_f$ of the largest critical avalanches, and we discuss the way to 
assess the validity of the predictions in 
computer simulations. We also discuss the RFIM on a Dyson hierarchical model 
with the same parameter 
$\sigma$. Based on the results given in the 
literature\cite{rodgers-bray,monthus11}, we conclude that this 
model displays a unique ``cuspless'' regime over the whole range of $\sigma$, 
with the avalanche-induced 
cuspy contributions being subdominant at the fixed point.

\section{Model and NP-FRG formalism}

We study the one-dimensional random-field Ising model with power-law decaying 
long-range ferromagnetic interactions.\cite{grin_76,Bray} 
It is described by the following Hamiltonian:
\begin{equation}
\begin{split}
 \label{Ham_LR}
 H=-\sum_{i,j}J_{ij} s_i s_j-\sum_i h_i s_i
\end{split}
\end{equation}
where $s_i=\pm 1$ are Ising spins placed on the vertices of a lattice, with a 
ferromagnetic pairwise interaction
\begin{equation}
 \label{JLR}
 J_{ij} \propto \vert x_i -x_j \vert ^{-(d+\sigma)}
\end{equation}
at long distance. The exponent $\sigma>0$ characterizes the long-range 
power-law decay of the 
interaction and $d$ is the spatial dimensionality. In the present case $d=1$. 
The disorder is introduced by the random fields $h_i$, which are independently 
distributed with a Gaussian distribution of width $\Delta_B$ 
around a zero mean: 
\begin{equation}
 \label{gauss}
 P(h_i)=\frac{1}{\sqrt{2\pi\Delta_B^2}}e^{-\frac{h_i^2}{2\Delta_B}} \, .
\end{equation}

To make use of the NP-FRG formalism and investigate the long-distance 
properties 
of the model near its critical 
point, it is more convenient reformulate the model in the field-theory setting. 
By using standard manipulations,  the Hamiltonian in Eq. (\ref{Ham_LR}) 
is replaced by the ``bare action'' for a scalar field $\varphi$:
\begin{equation}
\begin{split}
\label{action_bare}
&S[\varphi;h]=S_B[\varphi]-\int_x h(x)\varphi(x)\\&
S_B=\int_x\big [\frac{r}{2}\varphi(x)^2+\frac{u}{4!}\varphi(x)^4\big 
]+\frac{1}{2}\int_{x,y}\mathcal J(|x-y|)\varphi(x)\varphi(y)
\end{split}
\end{equation}
where $\int_x=\int d^dx$ and 
$h(x)$ is the continuous version of the random magnetic 
field with $\overline{h(x)}=0$ and 
$\overline{h(x)h(y)}=\Delta_B \delta^{(d)}(x-y)$ (where as usual the overline 
denotes the average over the quenched disorder). The interaction $\mathcal 
J(|x|)$ goes as 
$|x|^{-(d+\sigma)}$ at large distance and, accordingly, its Fourier transform 
behaves as
\begin{equation}
\label{eq_LRI}
\mathcal J( p)=\mathcal J \vert p\vert^{\sigma} + {\rm O}(p^2)
\end{equation}
when $p \rightarrow 0$.

From here, the NP-FRG equations can be derived along two alternative routes. In 
the first one,  
the system is considered directly at zero temperature and one builds a 
superfield formalism that can account for 
the fact that the equilibrium properties are given by the ground state of the 
model.\cite{tissier11,bacz_LR3d} 
In the second one, the system  is considered at finite temperature and one 
works with the 
Boltzmann weight.\cite{tarjus04,tissier06} The main advantage of the former is 
that it makes explicit 
the underlying supersymmetry that is responsible for the $d \rightarrow d-2$ 
dimensional-reduction property and allows one to study its spontaneous breaking 
along the NP-FRG flow.  
The latter one is however quite simpler to present and in the following we will 
follow this route. We stress that 
the two derivations lead to the \textit{same} exact NP-FRG equations for the 
zero-temperature fixed point 
controlling the critical behavior of the RFIM.

Due to the presence of quenched disorder the generating functional of the 
(connected) correlation functions, $\mathcal W_h[J]=\ln 
\int \mathcal D\varphi \,{\rm exp}(-S[\varphi;h]+\int_x J(x)\varphi(x))$, is 
random and can then be characterized by its cumulants. The latter are 
conveniently studied by considering copies or replicas of the system (see e.g. 
Refs. [\onlinecite{tarjus04,tissier11}]) which, differently from the standard 
replica 
trick,\cite{Par_book} are each coupled to a distinct external source. After 
averaging over the disorder, the resulting  ``multicopy'' generating functional 
$W[\{J_a\}]=\ln \overline{\prod_a {\rm exp}(\mathcal W_h[J_a])}$ is given by
\begin{equation}
\begin{split}
\label{bare_Zn}
&e^{W[\{J_a\}]}=\int \prod_a \mathcal{D}\varphi_a {\rm exp}\Big \{ \sum_a \Big 
(-S_B[\varphi_a] \\
 &+ \int_x J_a(x)\varphi_a(x)\Big )-\Delta_B 
\sum_{a,b}\int_x\varphi_a(x)\varphi_b(x)\Big\} \,.
\end{split}
\end{equation}
The cumulants are generated by expanding in increasing number of 
unrestricted (or free) sums over replicas:
\begin{equation}
W[\{J_a\}]=\sum_a W_{1}[J_a]+\frac{1}{2}\sum_{a,b}W_{2}[J_a,J_b]+\cdots,
\end{equation}
where $W_{k1}[J_a]=\overline{\mathcal W_{h,k}[J_a]}$ is the first cumulant, 
$W_{k2}[J_a,J_b]=\overline{\mathcal W_{h,k}[J_a]\mathcal W_{h,k}[J_b]}\vert_c$ 
the second cumulant, etc. 

The formulation of the NP-FRG proceeds by modifying the partition function in 
Eq. (\ref{bare_Zn}) with the introduction of an ``infrared regulator'' that 
suppresses the integration over the modes with momentum less than some cutoff 
$k$ and takes the form of a generalized ``mass'' (quadratic) term added to the 
bare action:\cite{Ber02}
\begin{equation}
\begin{split}
\label{regulator}
\Delta S_k[\{\varphi_a\}]=
\frac{1}{2}\sum_{a,b}\int_{x,y}\varphi_a(x)\mathcal R_{k,ab}(\vert 
x-y\vert)\varphi_b(y) \,,
\end{split}
\end{equation} 
where $\mathcal R_{k,ab}=\widehat{R}_k\delta_{ab} +\widetilde{R}_k$ with 
$\widehat{R}_k$ and 
$\widetilde{R}_k$ two functions enforcing an infrared cutoff on the 
fluctuations.\cite{tissier11} 
Through this procedure, one defines the  
generating functional of the (connected) correlation functions at scale $k$, 
$W_k[\{J_a\}]$.

In the NP-FRG approach, the central quantity is the ``effective average 
action'' $\Gamma_k$, which is the generating functional of the one-particle 
irreducible (1PI) correlation functions or vertices and is obtained from 
$W_k[\{J_a\}]$ 
through a (modified) Legendre transform:
\begin{equation}
 \Gamma_k[\{\phi_a\}]+\Delta S_k[\{\phi_a\}]=-W_k[\{J_a\}]+\sum_a\int_x 
J_a(x)\phi_a(x),
\end{equation}
where the field $\phi_a=\delta W_k/\delta J_a(x)$ is the average of the 
physical field in copy $a$. 

Similarly to $W_k[\{J_a\}]$,  $\Gamma_k[\{\phi_a\}]$ can be expanded in an 
increasing number of free replica sums,
\begin{equation}
\Gamma_k[\{\phi_a\}]=\sum_a 
\Gamma_{k1}[\phi_a]-\frac{1}{2}\sum_{a,b}\Gamma_{k2}[\phi_a,\phi_b]+\cdots \,,
\end{equation}
where the  $\Gamma_{kp}$'s are essentially the cumulants of the renormalized 
disorder.\cite{tarjus04,tissier11}

The $k$ dependence of the effective average action is governed by an exact 
renormalization-group (RG) equation:\cite{wetterich93,Ber02}
\begin{equation}
\label{ERGE}
\partial_t \Gamma_k[\{\phi_a\}]=\frac{1}{2}{\rm Tr}\widetilde{\partial}_t 
\ln(\Gamma^{(2)}_k[\{\phi_a\}]+\mathcal{R}_k)\, ,
\end{equation}
where $t=\ln(k/\Lambda)$, $\widetilde{\partial}_t$ is a symbolic notation 
indicating a derivative acting 
only on the $k$ dependence of the cutoff functions (\textit{i.e.},  
$\widetilde{\partial}_t \equiv 
\partial_t \widehat{R}_k\, \delta/\delta \widehat{R}_k + \partial_t 
\widetilde{R}_k \, \delta/\delta \widetilde{R}_k$), 
and $\Gamma^{(2)}_k$ is the second functional derivative of the effective 
average action with respect to 
the replica fields. (Generically, superscripts indicate functional 
differentiation with respect to the 
field arguments.) Finally, the trace involves summing over copy indices and 
integrating over spatial coordinates. 

The initial condition of the RG flow when $k=\Lambda$, the microscopic scale 
(\textit{e.g.} the inverse of the 
lattice spacing), is provided by the bare action and when $k \rightarrow 0$ one 
recovers the effective action of the 
full theory with all fluctuations accounted for. 

After expanding both sides of Eq. (\ref{ERGE}) in an increasing number of free 
replica sums, one obtains a hierarchy 
of exact RG flow equations for the cumulants of the renormalized disorder. For 
instance, the equations for 
first two cumulants read
\begin{equation}
\begin{split}
\label{eq_ERGE_Gamma1}
&\partial_t  \Gamma_{k1}[\phi_a]=\\&-\frac{1}{2} \tilde{\partial}_t {\rm tr}  
\Big \{ \ln \widehat{G}_k[\phi_a] + 
\widehat{G}_k[\phi_a] \Big (\Gamma_{k2}^{(11)}[\phi_a,\phi_a ] - 
\widetilde{R}_{k} \Big)\Big \} 
\end{split}
\end{equation}
\begin{equation}
\begin{split}
\label{eq_ERGE_Gamma2}
&\partial_t  \Gamma_{k2}[ \phi_a , \phi_b]= -\frac{1}{2}\tilde{\partial}_t {\rm 
tr} \bigg \{- \Gamma^{(101)}_{k3}[\phi_a,\phi_b,\phi_a]
\widehat{G}_k[\phi_a] + 
\\&
\Gamma^{(20)}_{k2}[\phi_a,\phi_b]\widetilde{G}_k[\phi_a,\phi_a] 
+\frac{1}{2}\Big 
(\Gamma_{k2}^{(11)}[\phi_a,\phi_b ] - \widetilde{R}_{k} \Big)
\widetilde{G}_k[\phi_a,\phi_b] \\&+ perm(a,b)\bigg \},
\end{split}
\end{equation}
where $perm(a,b)$ denotes the terms obtained by permuting the two indices $a$ 
and $b$ 
and the trace now only involves integrating over the spatial coordinates and 
the ``propagators'' 
$\widehat{G}_k$ and $\widetilde{G}_k$ are defined as
\begin{equation}
\begin{split}
\label{1_copy_prop}
\widehat{G}_{k;x_1x_2}[\phi_a]=\Big(\Gamma^{(2)}_{k1}[\phi_a]+\widehat{R}\Big)^{
-1}
\Big \vert_{x_1,x_2}
\end{split}
\end{equation}
\begin{equation}
\begin{split}
\label{2_copy_prop}
\widetilde{G}_{k;x_1x_2}[\phi_a,\phi_b]=&-\int_{x_3x_4}\widehat{G}_{k;x_1x_3}[
\phi_a]\Big (\Gamma^{(11)}_{k2;x_3x_4}[\phi_a,\phi_b]
\\& 
- \widetilde{R}_{k}(\vert x_3-x_4\vert)\Big )\widehat{G}_{k;x_4x_2}[\phi_b]\,.
\end{split}
\end{equation}

Up to now the RG equations are exact but they represent an infinite hierarchy 
of coupled functional equations and require approximations to be solved.

\section{Nonperturbative ansatz}
\label{sec:spec}

To truncate the exact hierarchy of  fuctional RG equations, a systematic 
nonperturbative approximation 
scheme has been proposed and successfully applied to the short-range 
RFIM\cite{tarjus04,tissier11} and the 
$3$-d RFIM with both long-range interactions and random-field 
correlations.\cite{bacz_LR3d} It 
consists in formulating an ansatz for the effective average action that relies 
on truncating the expansion in 
cumulants and approximating the spatial dependence of the fields through a 
truncated expansion in gradients (or 
in fractional Laplacians). 

We have adapted this approximation scheme to the present $1$-dimensional 
long-range model. The main specificities 
of this model compared to the random-field systems studied before through the 
NP-FRG are that

(i) the underlying supersymmetry that is responsible for the $d\rightarrow d-2$ 
dimensional-reduction property, 
\textit{i.e.} the superrotational invariance, is not present (and dimensional 
reduction is of course not an 
issue in $d=1$)

and

(ii) the small momentum dependence of the 2-point 1-copy 1PI vertex function 
acquires anomalous terms.

An efficient ansatz that can capture the long-distance physics including the 
influence of rare events such avalanches 
is then
\begin{equation}
\begin{split}
\label{eq_ansatz_gamma1}
\Gamma_{k1}[\phi_a]=&\int_x \Big \{U_k(\phi_a(x))+ \frac 12 
\mathcal J_k(\phi_a(x))\phi_a(x)(-\partial_{x}^2)^{\frac{\sigma}{2}} \phi_a(x)
\\& + \frac 12 Z_k(\phi_a(x))\phi_a(x)(-\partial_{x}^2)^{\frac{1+2\sigma}{2}} 
\phi_a(x)\Big \} \,,
\end{split}
\end{equation}
\begin{equation}
\begin{split}
\label{eq_ansatz_gamma2}
\Gamma_{k2}[\phi_a,\phi_b]=\int_x V_k(\phi_a(x),\phi_b(x))\,,
\end{split}
\end{equation}
\begin{equation}
\begin{split}
\label{eq_ansatz_gammap}
\Gamma_{kp\geq3}=0 \,
\end{split}
\end{equation} 
where $(-\partial_{x}^2)^{\alpha}$, with $\alpha$ a real number, denotes 
a fractional Laplacian: its Fourier transform generates a $(p^2)^{\alpha}$ term 
and for 
$\alpha=1$ it reduces to the usual Laplacian. The first of the 
fractional-Laplacian terms, 
$(-\partial_{x}^2)^{\frac{\sigma}{2}}$, directly stems from the long-range 
interaction [see Eq. (\ref{eq_LRI})]. 
The second fractional-Laplacian term, 
$(-\partial_{x}^2)^{\frac{1+2\sigma}{2}}$, is specific to the 
present $1$-dimensional long-range case. As will be shown further down, this 
term is generated under 
renormalization and for the range of $\sigma$ under consideration, with, 
$\sigma\leq\frac{1}{2}$, 
it dominates at long-distance the conventional $(-\partial_{x}^2)$ term
(said otherwise, $\vert p\vert^{1+2\sigma}$ is dominant in the infrared 
compared to $p^2$). 

When expressed at the level of the 2-point 1PI vertices, the above ansatz leads 
to
\begin{equation}
\begin{split}
\label{eq_ansatz_vertex1}
&\Gamma_{k1;x_1x_2}^{(2)}[\phi_a]=\Big 
\{U''_k(\phi_a(x_1))+\lambda_k(\phi_a(x_1))(-\partial_{x_1}^2)^{\frac{\sigma}{2}
} + \\&
\frac 12 \lambda'_k(\phi_a(x_1))(-\partial_{x_1}^2)^{\frac{\sigma}{2}} 
\phi_a(x_1)+Y_k(\phi_a(x_1))(-\partial_{x_1}^2)^{\frac{1+2\sigma}{2}}
\\& + \frac 12 Y'_k(\phi_a(x_1))(-\partial_{x_1}^2)^{\frac{1+2\sigma}{2}} 
\phi_a(x_1)\Big \}\delta(x_1-x_2) \,,
\end{split}
\end{equation}
\begin{equation}
\begin{split}
\label{eq_ansatz_vertex2}
\Gamma_{k2;x_1x_2}^{(11)}[\phi_a,\phi_b]=\Delta_k(\phi_a(x_1),
\phi_b(x_1))\delta(x_1-x_2)\,,
\end{split}
\end{equation}
where $\lambda_k(\phi_a)=\partial_{\phi_a}[\mathcal J_k(\phi_a)\phi_a]$,  
$Y_k(\phi_a)=\partial_{\phi_a}[Z_k(\phi_a)\phi_a]$, 
and $\Delta_k(\phi_a,\phi_b)=V_k^{(11)}(\phi_a,\phi_b)$. 

For the present $1$-d long-range model, we make a further simplifying step 
which is to set the 
cutoff function $\widetilde{R}_k$ to zero. The fluctuations are still 
suppressed in the infrared by the 
cutoff function $\widehat{R}_k$.\cite{tarjus04,tissier06} The role of 
$\widetilde{R}_k$ was to ensure 
that superrotational invariance 
is not explicitly broken at the level of the regulator. As there is no such 
supersymmetry in the present case 
(see above), it is not crucial to keep it and we find it more convenient to 
drop it.

The flow of the functions $U''_k$, $\lambda_k$ $Y_k$, and $\Delta_k$ appearing 
in 
Eqs. (\ref{eq_ansatz_vertex1},\ref{eq_ansatz_vertex2}) can be obtained by 
inserting the ansatz in the exact 
RG flow equations in Eqs. (\ref{eq_ERGE_Gamma1},\ref{eq_ERGE_Gamma2}), then 
considering uniform configurations of the replica fields and working in Fourier 
space. For instance, 
the flow equation for the 1-copy 2-point 1PI vertex becomes, in a graphical 
representation,
\begin{equation}
\begin{split}
\label{diagram_1cgamma2}
&\partial_t\Gamma^{(2)}_{k1}(p;\phi)=-\frac{1}{2}\tilde{\partial_t}\int_q \\& 
\Bigg(
\raisebox{-20pt}{\includegraphics[width=240pt,height=40pt]{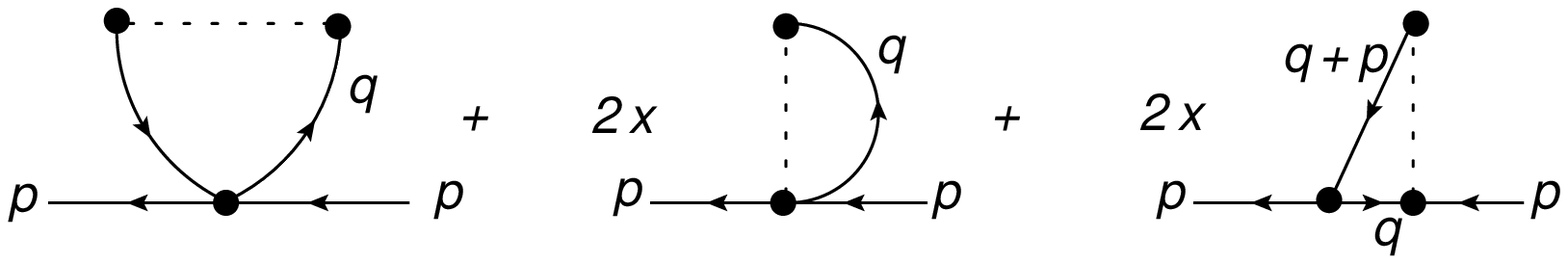}}
\\&
\raisebox{-15pt}{\includegraphics[width=180pt,height=36pt]{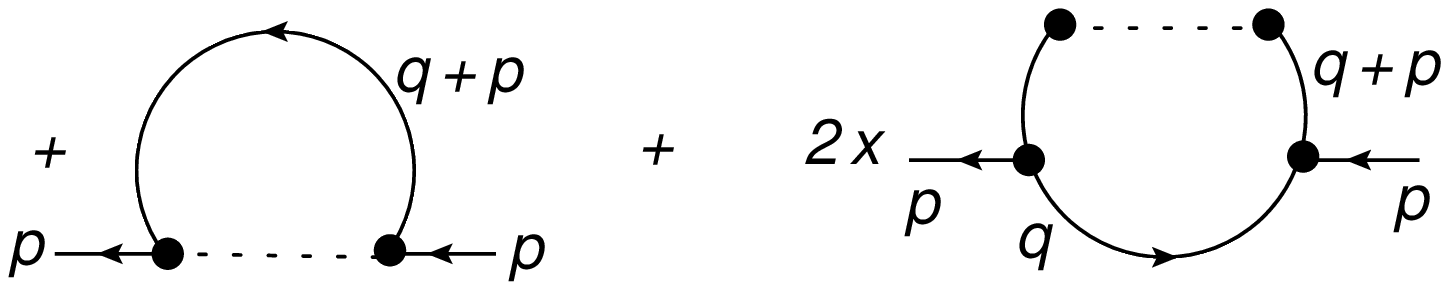}}
\Bigg),
\end{split}
\end{equation}
where lines denote the propagator $\widehat{G}_k$, dots the one-copy 1PI 
vertices and dots linked by dotted lines the two-copy 1PI vertices. The 
internal momentum 
is denoted by $q$ and the operator $\tilde{\partial}_t$ now acts only on 
$\widehat{G}_k$ 
through its dependence on $\widehat{R}_k$: $\tilde{\partial}_t 
\widehat{G}_k(q;\phi)=-\widehat{G}_k(q;\phi)^2\partial_t\widehat{R}_k(q)$.
To derive the above equation, we have used the fact that the expressions for 
the propagators $\widehat{G}_k$ 
and $\widetilde{G}_k$ in Eqs. (\ref{1_copy_prop},\ref{2_copy_prop}) can be 
simplified for uniform fields, with 
\begin{equation}
\label{FT2exprssn}
  \widetilde{G}_k(p;\phi_a,\phi_b)= \widehat{G}_k(p;\phi_a) 
\widehat{G}_k(p;\phi_b)\Delta_k(\phi_a,\phi_b)
\end{equation}
and
\begin{equation}
\label{FT1exprssn}  
\widehat{G}_k(p;\phi)=\frac{1}{\lambda_k(\phi)\vert 
p\vert^{\sigma}+Y_k(\phi)\vert p\vert^{
1+2\sigma}+\widehat{R}_k( p)+U''_k(\phi)}\,.
\end{equation}

The RG flow of $U''_k$ is then obtained from Eq. (\ref{diagram_1cgamma2}) when 
$p= 0$ and 
those of $\lambda_k$ and $Y_k$ by expanding the right-hand side of Eq. 
(\ref{diagram_1cgamma2}) 
in small $p$ and identifying the anomalous dependence in $\vert 
p\vert^{\sigma}$ and $\vert p\vert^{1+2\sigma}$. 
A similar procedure with the 2-copy 2-point 1PI vertex for $p=0$ allows one to 
derive the flow 
of $\Delta_k(\phi_a,\phi_b)$.

We find that the flow equation for $\lambda_k(\phi)$ is such that if 
$\lambda_k$ is independent 
of the field in the initial condition [see Eq. (\ref{eq_LRI})], it does not 
flow and remains equal to its 
bare value for all RG times. So, without loss of generality, we can set  
$\lambda_k= 1$. On the other hand, the flow equations for $U''_k(\phi)$ and 
$\Delta_k(\phi_a,\phi_b)$ 
are given in Appendix \ref{app:u''delta}.

Finally, the derivation of the flow equation for the function $Y_k(\phi)$ needs 
some special care 
and demonstrates why the small-momentum dependence is described by a term 
$\propto \vert p\vert^{1+2\sigma}$ (in 
the range $\sigma <\frac{1}{2}$). The key point is that even if one starts  
the RG flow with an initial condition where such a term is absent, it is 
generated along the flow. 

To obtain the flow equation for $Y_k$ one needs to isolate all of the terms 
contributing to the order $\vert p\vert^{1+2\sigma}$ when the right-hand side 
of Eq. (\ref{diagram_1cgamma2}) 
is expanded in small $p$. There are two types of terms:
\begin{itemize}
 \item[(i)] \textit{The vertex terms} - obtained by collecting the $\vert 
p\vert^{1+2\sigma}$
dependence from the 1PI vertices and dropping the $p$ dependence of the 
propagators. These terms are proportional to $Y_k$ and its derivatives.
 \item[(ii)] \textit{The anomalous terms} - produced by the singular momentum 
dependence in 
$\vert p\vert^{\sigma}$ present in the propagators appearing in the 1-loop 
integrals. 
This point is further explained for a toy model in Appendix \ref{toy_p}. 
These terms generate a contribution to the effective average action $\propto 
\vert p\vert^{1+2\sigma}$ even if it is not present in the initial condition of 
the flow. 
\end{itemize}

The expression for the flow equation of the function $Y_k$ can therefore be 
written as the 
sum of a vertex contributions $\beta_{Y,ve}$ and an anomalous $\beta_{Y,an}$, 
$ \partial_t Y_k(\phi)=\beta_{Y,ve}(\phi)+\beta_{Y,an}(\phi)$. 
The derivation of $\beta_{Y,ve}$ is relatively straightforward and is not 
detailed 
whereas that of $\beta_{Y,an}$ is sketched in Appendix \ref{app:betaY_an}. We 
give 
here the final expressions: 
\begin{equation}
\begin{split}
  \label{betY1}
&\beta_{Y,ve}(\phi)=\\&
\int_0^{+\infty} \frac{dq}{\pi} 
\partial_t\widehat{R}_k(q)\widehat{G}_k(q;\phi)^3\Big 
\{\Big(\Delta^{(1,0)}_k(\phi,\phi)
+  \Delta^{(0,1)}_k(\phi,\phi)\Big)  \\& 
\times 
Y'_k(\phi)+\Delta_k(\phi,\phi)\Big(Y''_k(\phi)-3\widehat{G}
_k(q;\phi)Y'_k(\phi)\big [U'''_k(\phi) \\& 
+ q^{1+2\sigma}Y'_k(\phi)\big ]\Big)\Big \}\,,
\end{split}
\end{equation}
\begin{equation}
\begin{split}
  \label{betY2}
&\beta_{Y,an}(\phi)=\\& 
\frac{2^{2\sigma}\sigma^2\Gamma(\frac{1}{2}-\sigma)}{\sqrt{
\pi}\Gamma(2-\sigma)}\,
\partial_t\widehat{R}_k(0)  \widehat{G}_k(0;\phi)^5 U'''_k(\phi)  \Big(5 \, 
\widehat{G}_k(0;\phi)
\\&
\times U'''_k(\phi) 
 \Delta_k(\phi,\phi) -2\big 
[\Delta^{(1,0)}_k(\phi,\phi)+\Delta^{(0,1)}_k(\phi,\phi)\big ]\Big)\,.
\end{split}
\end{equation}
Note that the expression of $\beta_{Y,an}(\phi)$ is nonzero even when $Y_k=0$.

\section{Fixed-point equations}
\label{sec-FP}

To describe the long-distance physics near the critical point and search for 
fixed points of the 
NP-FRG equations we first need to cast the latter in a dimensionless form. As 
the critical physics 
is related to a ``zero-temperature'' fixed point, one needs to introduce a 
renormalized temperature 
$T_k$ and an associated critical exponent $\theta >0$, such that $T_k\propto 
k^{\theta}$.\cite{tarjus04,tissier06} The renormalized 
cumulants then scale as
$\Gamma_{k1}\sim T_k^{-1}$, $\Gamma_{k2}\sim T_k^{-2}$ and the replica fields 
as 
$\phi_a \sim k^{(d-2+\eta)/2} T_k^{-1/2}$. 
Since $\lambda_k$, the term in $\vert p\vert^{\sigma}$ in the 2-point 1PI 
vertex is not renormalized 
(see above), one immediately derives that the anomalous dimension $\eta$ is 
always given by 
$\eta=2-\sigma$.

One can thus introduce dimensionless quantities as follows (recall that $d=1$):
\begin{equation}
\begin{split}
&\phi_a =  k^{(1-\sigma)/2}T_k^{-1/2}\varphi_a \sim 
k^{\frac{-3+\overline{\eta}}{2}}  \\&
 U''_k(\phi) = k^{\sigma} u''_k(\varphi) \\&
 Y_k (\phi)= k^{-(1+\sigma)} y_k(\varphi)\\&
 \Delta_k (\phi_a,\phi_b) =  k^{-\sigma} T_k^{-1} \delta_k(\varphi_a,\varphi_b) 
\sim k^{-(2\eta-\bar\eta)}\,,
\end{split}
\end{equation}
where the additional anomalous dimension $\bar\eta$ is related to the 
temperature exponent $\theta$ and to $\eta$ through 
$\theta=2+\eta-\bar\eta$.

The cutoff function $\widehat{R}_k$ can also be put in a dimensionless form, 
$\widehat{R}_k(q)=k^{\sigma}s(q^2/k^2)$, 
and we have used $s(x^2)=(a+bx^2+cx^4)e^{-x^2}$ with the parameters $a,b,c$ 
optimized through stability 
considerations and varied to provide error bars on the 
results.\cite{litim_opt,canet_opt,pawlowski_opt}

We thus have to solve three coupled dimensionless flow equations, which can be 
symbolically written as
\begin{equation}
\begin{split}
  \label{eq_dimensionless}
&\partial_t u''_k(\varphi)=\widetilde{\beta}_{u''}(\varphi) \\&
\partial_ty_k(\varphi)=\widetilde\beta_{y}(\varphi)\\&
\partial_t\delta_k(\varphi_{a},\varphi_{b})=\widetilde\beta_{\delta}(\varphi_{a}
,\varphi_{b})\,.
\end{split}
\end{equation}
The (running) exponent $\bar{\eta}_k$ can be calculated from the flow 
equation for $\delta_k$ by imposing that $\partial_t \delta_k=0$ for a given  
arbitrary value of the fields. This directly gives an expression for 
$2\eta-\bar\eta_k$. The result is 
not sensitive to the field values and we have chosen 
the point $\varphi_a=\varphi_b=0$: then, without loss of generality, 
we impose $\delta_k(0,0)=\delta_{\Lambda}\equiv 1$.

We have solved numerically the fixed-point equations, obtained by setting to 
zero the left-hand sides of 
Eqs. (\ref{eq_dimensionless}) for $\frac{1}{3}\leq \sigma<\frac{1}{2}$. To do 
so, we have discretized the fields on a grid 
and used a variation of the Newton-Raphson method.\cite{newt_rasp}  

\section{Results}
\label{sec-results}

\begin{figure}[h!]
  \begin{center}
    \includegraphics[width=250pt,height=200pt]{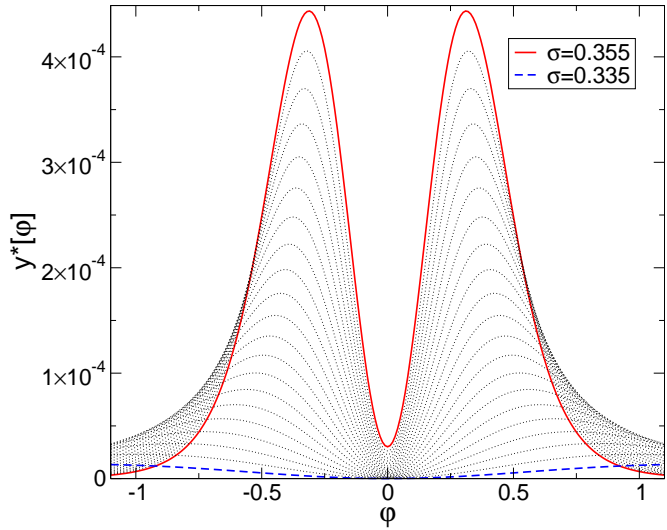}  
    \caption{Fixed-point solutions for the function $y^{*}(\varphi)$ for 
several 
values of $\sigma$ close to $\sigma_G=1/3$. The system at $\sigma_G=1/3$ is 
equivalent to a system at the upper critical dimension: the fixed point is 
Gaussian (in particular,
 $y¬*=0$). Another fixed point emerges for $\sigma>\sigma_G$.}\label{fig3}
  \end{center}
\end{figure}

For illustration, we first display in Fig. \ref{fig3} the fixed-point function  
$y^{*}(\varphi)$ for a range of values of $\sigma$ near $\sigma_G=1/3$. The 
system at $\sigma_G=1/3$ is 
equivalent to a system at the upper critical dimension: the fixed point is 
Gaussian (in particular, 
$y¬*=0$) and the critical exponents for $\sigma<\sigma_G$ are the classical 
(mean-field) 
ones for a long-range system\cite{Bray} (see also below). Above $\sigma_G$ the 
Gaussian fixed point 
is unstable due to the $\phi^4$ term in the potential $U_k(\phi)$ and another 
fixed point emerges. 

For $1/3<\sigma<1/2$, we find two different regimes for the critical behavior, 
separated by a 
critical value $\sigma_c\approx 0.379$. What distinguishes these regimes is the 
presence or the absence of 
a nonanalyticity in the form of a linear cusp in the (dimensionless) 
renormalized second cumulant of the 
random field at the fixed point, $\delta^*(\varphi_{a},\varphi_{b})$, when 
$\varphi_{b} \rightarrow \varphi_{a}$:  
a cusp is present when $\sigma > \sigma_c$ but is absent when  $\sigma < 
\sigma_c$. 
The existence of two such regimes separated by a critical value, 
has been found in the short-range RFIM,\cite{tarjus04,tissier06,tissier11} in 
the RFO(N)M\cite{tissier06} and in the long-range 
RFIM in $d=3$ as well.\cite{bacz_LR3d} In these cases, the cuspless fixed point 
is associated with a critical 
behavior satisfying the dimensional-reduction property whereas the presence of 
a cusp breaks the dimensional reduction 
and the underlying supersymmetry. In the present $1$-dimensional model, there 
is no dimensional reduction and 
no underlying supersymmetry whatsoever. The presence of a cusp in the 
dimensionless cumulant of the disorder 
has an effect on the values of the critical exponents, but not as dramatic as 
in the cases involving dimensional 
reduction and its breakdown.

\begin{figure}[h!]
    \begin{center}
      \includegraphics[width=250pt,height=200pt]{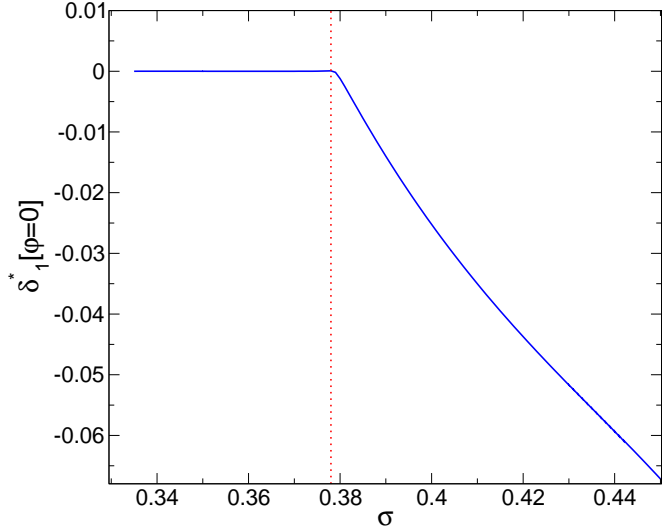}  
    \caption{Fixed-point value of the amplitude of the cusp, 
$\delta_1^{*}(\varphi=0)$, 
versus $\sigma$. Below $\sigma_c\approx 0.379$ (denoted by the dotted vertical 
line), the fixed-point solution is cuspless around $\delta\varphi=0$; a cusp 
appears and its magnitude further grows above $\sigma_c$.}\label{fig4}
  \end{center}
 \end{figure}

This nonanalytical dependence in the dimensionless fields is more conveniently 
studied by introducing
\begin{equation}
\begin{split}
& \varphi=\frac{\varphi_a+\varphi_b}{2} \\&
 \delta\varphi=\frac{\varphi_b-\varphi_a}{2} \,.
\end{split}
\end{equation}
The cumulant $\delta^{*}$ is an even function of $\varphi$ and 
$\delta\varphi$ separately. The presence of a cusp then means that
\begin{equation}
\label{eq_cusp_delta}
\delta^{*}(\varphi,\delta\varphi)=\delta^{*}_0(\varphi)+\delta^{*}
_1(\varphi)|\delta\varphi|+\frac{1}{2}\delta^{*}_2(\varphi)\delta\varphi^2 
+\cdots
\end{equation}
with $\delta^{*}_1(\varphi) \neq 0$. We show in Fig. \ref{fig4} 
$\delta^{*}_1(\varphi=0)$ versus $\sigma$. It is 
equal to zero below $\sigma_c \approx 0.379$, indicating that the fixed point 
is ``cuspless'', and it becomes 
nonzero above $\sigma_c$, signaling a ``cuspy'' fixed point.

For $\sigma<\sigma_c$, the cuspless 
fixed point and its vicinity are described by a renormalized second cumulant 
that behaves as
\begin{equation}
\label{eq_analytic}
\delta_k(\varphi,\delta\varphi)=\delta_{k,0}(\varphi)+\frac{1}{2}\delta_{k,2}
(\varphi)\delta\varphi^2 
+\cdots
\end{equation}
at small $\delta\varphi$. When inserting Eq. (\ref{eq_analytic}) into the RG 
flow equations,
 Eqs. (\ref{eq_dimensionless}), it is easily 
realized that one obtains a closed system of 3 coupled equations for 
$u''_k(\varphi)$, $y_k(\varphi)$ and 
$\delta_{k,0}(\varphi)$ that can be solved independently of the full dependence 
of $\delta_k(\varphi,\delta\varphi)$. 
A flow equation for $\delta_{k,2}(\varphi)$ is further obtained with a beta 
function that depends on 
$u''_k(\varphi)$, $y_k(\varphi)$, $\delta_{k,0}(\varphi)$ and 
$\delta_{k,2}(\varphi)$ only. The derivation is 
straightforward but cumbersome and the resulting equation is not shown here 
(see Refs. [\onlinecite{tarj_avlnch,balog_FP}] 
for a similar derivation). The function $\delta_{k,2}(\varphi)$ blows up at a 
finite scale $k_L$ for 
$\sigma>\sigma_c$ and the fixed-point function is then given by 
Eq.~(\ref{eq_cusp_delta}). On the other hand it 
reaches a finite fixed-point function for $\sigma<\sigma_c$. This allows us to 
find a precise estimate of 
$\sigma_c=0.379\pm0.001$ within the present approximation.

Before discussing in more detail the significance and the physics of the two 
regimes, it is instructive to 
characterize how the fixed point changes from ``cuspless'' to ``cuspy'' by 
considering the stability of the fixed point with respect to a ``cuspy'' perturbation, \textit{i.e.} a 
perturbation $\propto \vert \delta\varphi\vert$ at small $\delta\varphi$. To compute the 
eigenvalue $\lambda$ associated with such a ``cuspy'' perturbation, one then 
has to consider the vicinity of the fixed point with 
$\delta_k(\varphi,\delta \varphi)\simeq \delta^*(\varphi\delta \varphi) 
+k^{\lambda} f_{\lambda}(\varphi,\delta \varphi)$ with 
$f_{\lambda}(\varphi,\delta \varphi) \simeq \vert \delta \varphi \vert 
f_{\lambda}(\varphi)$ 
when $\delta \varphi \rightarrow 0$. By linearizing the flow equation for 
$\delta_k$ around $\delta^*$ and expanding around 
$\delta \varphi= 0$ one easily derives the eigenvalue equation for $ 
f_{\lambda}(\varphi)$, which depends on 
$u''^*(\varphi)$, $y^*(\varphi)$, $\delta_{0}^*(\varphi)$ and 
$\delta_{2}^*(\varphi)$ only. The details are similar to  
those given in the supplementary material of Ref. [\onlinecite{tarj_avlnch}] 
and the resulting equation is not reproduced here.

\begin{figure}[h!]
    \begin{center}
      \includegraphics[width=250pt,height=200pt]{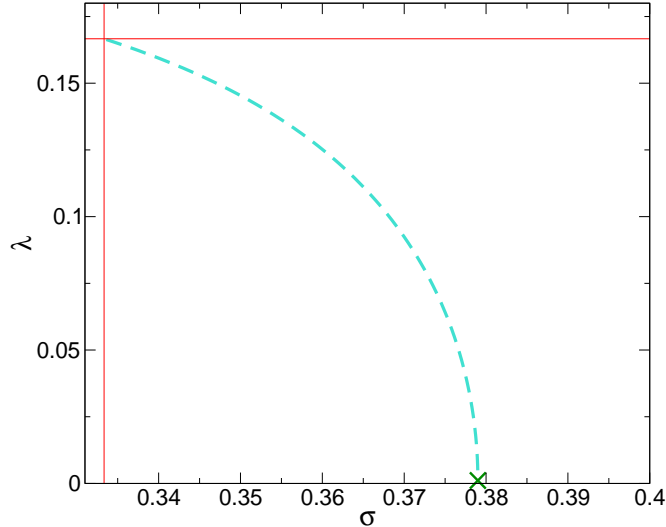}  
    \caption{The eigenvalue $\lambda$ associated with a cuspy perturbation 
versus $\sigma$. The dashed line is the calculation around the cuspless fixed 
point for $\sigma \leq \sigma_c\approx 0.379$: $\lambda$ decreases from $1/6$ 
when $\sigma=\sigma_G=1/3$ to a very small but nonzero value in $\sigma_c$; 
above $\sigma_c$ the fixed point is characterized by a cusp and $\lambda$ is 
strictly zero.}\label{fig5}
  \end{center}
 \end{figure}

We show in Figure \ref{fig5} the evolution of $\lambda$ with $\sigma$. It is 
positive for $\sigma<\sigma_c$, thereby 
indicating that the cuspy perturbation is irrelevant. For 
$\sigma=\sigma_G=1/3$, 
\textit{i.e.} around the Gaussian fixed point, $\lambda$ can be exactly 
computed and is equal to $1/6$. The 
eigenvalue then decreases as $\sigma$ increases. The $1$-loop perturbative 
result in $\sigma=\sigma_G+ \epsilon$, 
which is reproduced by the present nonperturbative ansatz, is found to be 
$\lambda=1/6 - \epsilon$ 
(see Appendix \ref{NPRGtoPERT}). When $\sigma=\sigma_c^-$, we obtain a very 
small yet strictly positive value, $\lambda_{c-}=0.0011\pm0.0001$. (This is a robust result as the value is 
always found strictly positive when varying the parameters of the cutoff function.) Above $\sigma_c$, 
the fixed point is cuspy, and the eigenvalue $\lambda$ can therefore be taken as being zero. In consequence, 
$\lambda$ is discontinuous in $\sigma_c$, which, as explained in Ref. [\onlinecite{balog_FP}], is 
the signature that the cuspy fixed emerges continuously from the 
cuspless one through a boundary-layer mechanism.

We now turn to the values of the critical exponents. In addition to $\eta$, 
which is fixed to $2-\sigma$ due to 
the long-range nature of the interactions, the critical behavior of the system 
is characterized by the correlation-length exponent $\nu$ and the exponent $2\eta-\bar{\eta}$ that 
characterizes the scaling behavior of the 
renormalized cumulant of the random field. From these exponents, one can deduce 
the other criticalexponents through scaling relations (which are exactly satisfied by the NP-FRG).

\begin{figure}[h!]
    \begin{center}
      \includegraphics[width=250pt,height=200pt]{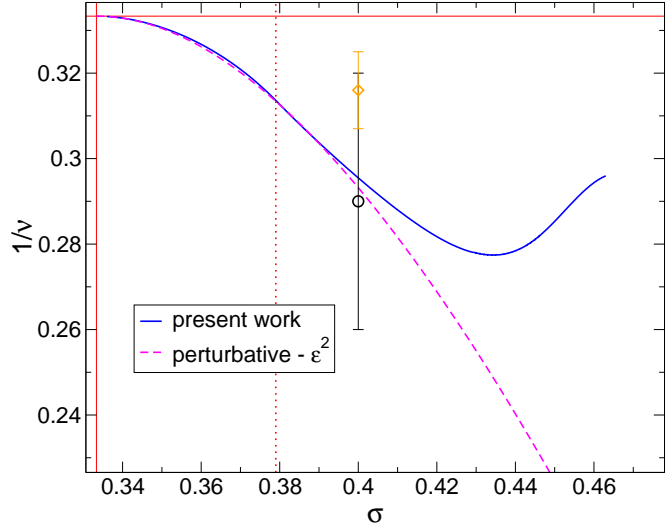}  
    \caption{Inverse of the correlation-length exponent $\nu$ versus $\sigma$ 
as 
obtained from the present NP-FRG theory (full line). The dashed line is the 
perturbative result at order $\epsilon^2$ around 
$\sigma_G=1/3$.\cite{young_77,Bray} The symbols with the error bars are results 
from lattice simulations: Ref. [\onlinecite{dewenter}] (circle), Ref. 
[\onlinecite{leuzzi}] (diamond). 
The dotted vertical line marks 
the value $\sigma_c \approx 0.379$ separating the ``cuspless'' and the 
``cuspy'' regimes.}
\label{fig6}
  \end{center}
 \end{figure}

\begin{figure}[h!]
    \begin{center}
      \includegraphics[width=250pt,height=200pt]{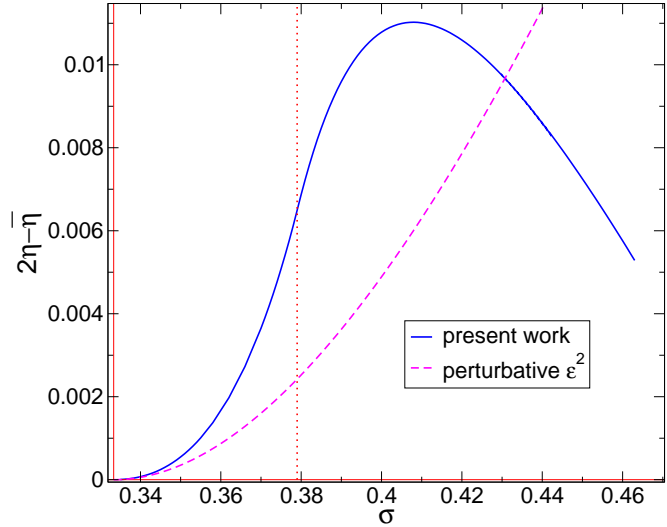}  
    \caption{$2\eta-\bar\eta$ versus $\sigma$ as obtained from the present 
NP-FRG theory (full line). The dashed line is the perturbative result to order 
$\epsilon^2$ around $\sigma_G=1/3$.\cite{Bray} The dotted vertical line marks 
the value $\sigma_c \approx 0.379$ 
separating the ``cuspless'' and the ``cuspy'' regimes.}
\label{fig7}
  \end{center}
 \end{figure}

The results for $1/\nu$ and $2\eta-\bar{\eta}$ versus $\sigma$ are displayed in 
Figs. \ref{fig6} and \ref{fig7}, respectively. Both $1/\nu$ and $2\eta-\bar{\eta}$ have a 
nonmonotonic behavior with $\sigma$ 
but this is not related to the change of regime at $\sigma_c$. At and around 
$\sigma_c$ the variation of the two exponents is smooth, as expected from the boundary-layer 
mechanism.\cite{balog_FP} It is 
noteworthy that the value of $2\eta-\bar{\eta}$ is small over the whole range 
of $\sigma$, being at most of the order of $.01$.

As already mentioned, the present NP-FRG equations exactly reproduce the known 
$1$-loop perturbative results for 
$\sigma=\sigma_G+\epsilon$: $1/\nu=1/3 + {\rm O}(\epsilon^2)$ and 
$2\eta-\bar{\eta}={\rm O}(\epsilon^2)$ 
(see Appendix \ref{NPRGtoPERT}). They are not exact at order $\epsilon^2$ but 
the coefficients of the $\epsilon^2$ terms are found to be close to the exact 
ones:\cite{young_77,Bray} $-10.61$ for  
$1/\nu$ and $1.498$ for $2\eta-\bar{\eta}$ to be compared to the exact 
$-10.8984\cdots$ and $1.32498\cdots$. The NP-FRG and the 
perturbative result at order $\epsilon^2$ stay close up to $\sigma \sim0.4$ for 
$1/\nu$ but strongly differ even in the 
vicinity of $\sigma_G=1/3$, which seems to indicate in this case large 
higher-order corrections. None of this however 
appears to be related to the change of regime at 
$\sigma_c$. Note finally that the result for $1/\nu$ at $\sigma=0.4$, 
\textit{i.e.} in the ``cuspy'' regime, is 
in very good agreement with the value recently obtained  from numerical 
ground-state determination by  
Dewenter and Hartmann\cite{dewenter} (a little less with the result of Ref. 
[\onlinecite{leuzzi}]).

We have been able to follow the critical fixed point up to $\sigma \approx 
0.46$, after which the numerical instabilities proliferate. The reason is that 
$\sigma$ approaches the value of 
$1/2$, which is the analog of a lower critical dimension below which there is 
no transition. It is then 
expected that the low-momentum dependence of the propagators is no longer well 
described by only taking into 
account the lowest-order terms of the derivative expansion as done in the 
ansatz used in this work. 

\section{Discussion}

The existence of two different regimes in the critical behavior of the RFIM can 
be attributed to the role of 
``avalanches'', which are collective phenomena present at zero temperature. In 
equilibrium, such ``static'' avalanches describe the discontinuous change 
in the ground state of the system at values of the external source 
that are sample-dependent (for an illustration see the figures in Refs. 
[\onlinecite{wu-machta05,liu-dahmen07}]). At the critical point, 
the avalanches take place at all scales and their size distribution follows a 
power law with a nontrivial exponent $\tau$. 

It was shown in Ref. [\onlinecite{tarj_avlnch}] for the RFIM (and in Refs. 
[\onlinecite{BBM96,aval_FRG}] for 
interfaces in a random environment) that the presence of avalanches necessarily 
induces a cusp in the functional  dependence of the renormalized cumulants of the disorder 
and in properly chosen correlation functions. 
Actually, such a cuspy dependence is already present in the $d=0$ 
RFIM.\cite{tarj_avlnch} As the critical behavior 
is affected by  a cusp in the \textit{dimensionless} 
quantities,\cite{tarjus04,tissier06,tissier11} the 
question is then whether the amplitude of the cusp is, or not, an irrelevant 
contribution at the fixed point. 

In physical terms, the two regimes therefore correspond to two different 
situations where the scale of the largest typical 
avalanches at criticality, which in a system of linear size $L$ goes as 
$L^{d_f}$ with $d_f \leq d$, is equal to 
the scale of the total magnetization, \textit{i.e.} as 
$L^{d-(d-4+\bar\eta)/2}$, or is subdominant.  This is precisely 
characterized by the eigenvalue $\lambda$ calculated in the previous section. 
In the regime where the fixed 
point is cuspless, $d_f=d-(d-4+\bar\eta)/2-\lambda$ with $\lambda >0$ 
whereas $d_f=d-(d-4+\bar\eta)/2$ in the regime where the fixed point is cuspy. 

The main interest of the $1$-dimensional long-range RFIM is that it can be 
studied by computer simulations for large system sizes. 
Efficient algorithms exist to determine the ground state in a polynomial time 
and one can further reduce the computational 
cost by considering  a diluted (Levy) lattice which is expected to be in the 
same universality class.\cite{leuzzi,dewenter} 
Sizes up to $L=256000$ in Ref. [\onlinecite{leuzzi}] and $L=524288$  in Ref. 
[\onlinecite{dewenter}] have thus been investigated.

As discussed before, the critical exponents characterizing the leading scaling 
behavior do not show any significant 
change of dependence with $\sigma$ around the critical value $\sigma_c$ (they 
are predicted to be continuous in $\sigma_c$). 
To distinguish the two regimes it seems therefore preferable to investigate the 
avalanche distribution at $T=0$ and try 
to extract the fractal dimension $d_f$ of the largest avalanches. This can be 
obtained through a finite-size scaling of the 
distribution of the avalanches or of its first moments. For instance, the ratio 
of the second to the first moment of the 
avalanche size, $<S^2>/<S>$, should scale as $L^{d_f}$ for $L$ large enough. 
Having extracted $d_f$ and $2\eta -\bar\eta$, one can obtain the eigenvalue $\lambda$ from the relation
\begin{equation}
\label{eq_lambda}
\lambda= \frac{d+4-\bar\eta}{2}-d_f=\frac 12 +\sigma +\frac{2\eta-\bar\eta}{2} -d_f
\end{equation}
where $2\eta -\bar\eta$ is anyhow expected to be very small. If $\lambda >0$, 
one is in the cuspless regime 
and if $\lambda=0$ in the cuspy one. By studying values of $\sigma$ near 
$\sigma_G=1/3$ where $\lambda$ 
should be $\approx 1/6$ and values of $\sigma \gtrsim 0.4$ it should be 
possible to distinguish the two regimes in computer simulations. 

In addition, one could also consider the corrections to scaling. In the 
vicinity of the critical value $\sigma_c\approx 0.379$, these corrections should be dominated 
by the lowest (positive) eigenvalue in the $Z_2$-symmetric subspace of 
perturbations around the fixed point. For $0.372\lesssim \sigma \leq \sigma_c$, 
this eigenvalue is equal to $\lambda$ and is therefore expected to be very small (with 
$\lambda_c^-\approx 0.0011$). Right above $\sigma_c$, 
the smallest irrelevant eigenvalue is also associated with 
a cuspy perturbation and it emerges from a value $\lambda_c^+\lesssim \lambda_c^-$ 
at $\sigma_c^+$.\cite{balog_FP} The exponent describing the 
main correction to scaling should therefore display a minimum as a function of 
$\sigma$ with a value near zero for $\sigma=\sigma_c\approx 0.379$.

Finally, we conclude by discussing the RFIM on the Dyson hierarchical lattice. 
This lattice mimics the behavior of the 
$1$-dimensional chain with long-range interactions specified by the same 
parameter $\sigma$.  Its critical behavior 
was studied by Rodgers and Bray\cite{rodgers-bray} and its avalanche 
distribution was characterized in detail more recently by Monthus and Garel.\cite{monthus11}

It was shown in Ref. [\onlinecite{rodgers-bray}] that in this case the 
temperature exponent is always given by $\theta=\sigma$, 
which implies, together with the result $\eta=2-\sigma$, that 
$2\eta-\bar\eta=0$. In addition, it was found in 
Ref. [\onlinecite{monthus11}] that the fractal dimension of the largest 
avalanches at criticality is $d_f=2\sigma$. 
These two results, $2\eta-\bar\eta=0$ and $d_f=2\sigma$, when inserted into Eq. 
(\ref{eq_lambda}) predict that 
$\lambda=1/2-\sigma$. As a consequence, $\lambda$ decreases from $1/6$ to $0$ 
as $\sigma$ decreases from $1/3$ to $1/2$ and is always strictly positive, except at the 
analog of the lower critical dimension. 
In consequence, the RFIM on the Dyson hierarchical level appears to display a 
unique regime over the whole range 
$1/3 \leq \sigma <1/2$: avalanches are present at all scales but their scaling 
dimension is never large enough to 
induce a cuspy fixed point. This is a notable difference with the long-range 
RFIM on the standard $1$-d chain.

To summarize: The NP-FRG predicts that the critical behavior of the RFIM 
generically shows two distinct 
regimes separated by a nontrivial critical value of the dimension, the number 
of components or the power-law 
exponent of the long-range interactions, when present. These regimes are 
associated with properties of the 
large-scale avalanches at zero temperature. We suggest that investigating 
through numerical simulations the 
avalanche characteristics in the $1$-dimensional long-range RFIM should provide 
a direct check of the prediction.

\appendix

\section{The flow equations for $U''_k(\phi)$ and $\Delta_k(\phi_a,\phi_b)$}
\label{app:u''delta}

Starting from Eq.~(\ref{diagram_1cgamma2}) and setting the external 
momentum to $0$ one easily obtains the flow equation for $U''_k$, 
$\partial_t U''_k(\phi)=\beta_{U''}(\phi)$, with
   \begin{equation}
\begin{split}
\label{eq_beta_u''}
&\beta_{U''}=-\frac{1}{2}\int_{0}^{\infty} \frac{dq}{\pi} \partial_t 
\widehat{R}(q)\widehat{G}_kq;\phi)^2 \Big \{\Delta^{(2,0)}_k(\phi,\phi)
+ \\& 
\Delta^{(0,2)}_k(\phi,\phi) +2\Delta^{(1,1)}_k(\phi,\phi) 
+ 2\widehat{G}_k(q;\phi)\Big[-2 \times \\&
\big [\Delta^{(1,0)}_k(\phi,\phi)+\Delta^{(0,1)}_k(\phi,\phi)\big ]  \big 
[U'''_k(\phi)+q^{1+2\sigma}Y'_k(\phi) \big ]
\\& 
-\Delta_k(\phi,\phi) \Big(U^{(4)}_k(\phi)+q^{1+2\sigma}Y''_k(\phi) 
-3\widehat{G}_k(q;\phi) \times 
\\&
\big [U'''_k(\phi)+  q^{1+2\sigma} Y'_k(\phi_a)\big ]^2\Big)\Big]\Big \}
\end{split}
\end{equation}

To obtain the expression for the flow equation of $\Delta_k$, $\partial_t 
\Delta_k(\phi_a,\phi_b)=\beta_{\Delta}(\phi_a,\phi_b)$, we derive  Eq.~(\ref{eq_ERGE_Gamma2}) with 
respect to $\phi_a$ and $\phi_b$, 
insert the ansatz for the effective average action and express the output in 
terms of the Fourier 
transformed quantities. The flow of $\Delta_k$ is then given by setting the 
external momentum $p$ 
to zero, which leads to the following graphical expression: 
\begin{equation}
\begin{split}
\label{diagram_2cgamma2}
&\partial_t\Delta_{k}(\phi_a,\phi_b)=-\frac{1}{2}\tilde{\partial_t}\int_q
\\&
\Bigg(\raisebox{-20pt}{\includegraphics[width=240pt,height=40pt]{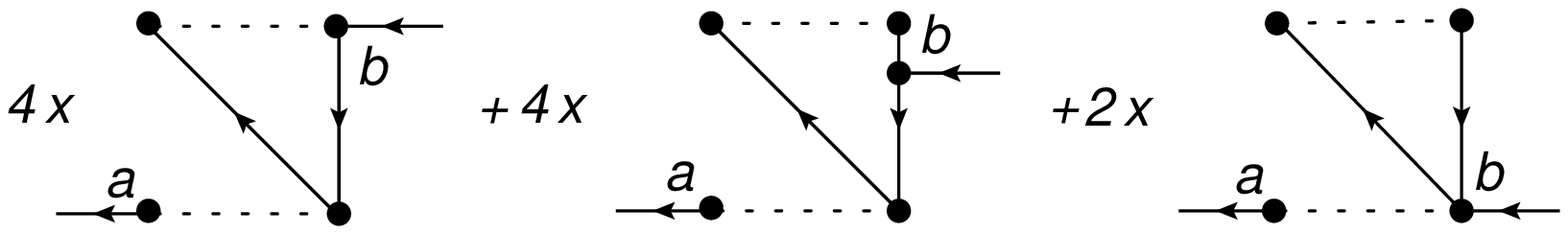}}
\\&
\raisebox{-15pt}{\includegraphics[width=240pt,height=40pt]{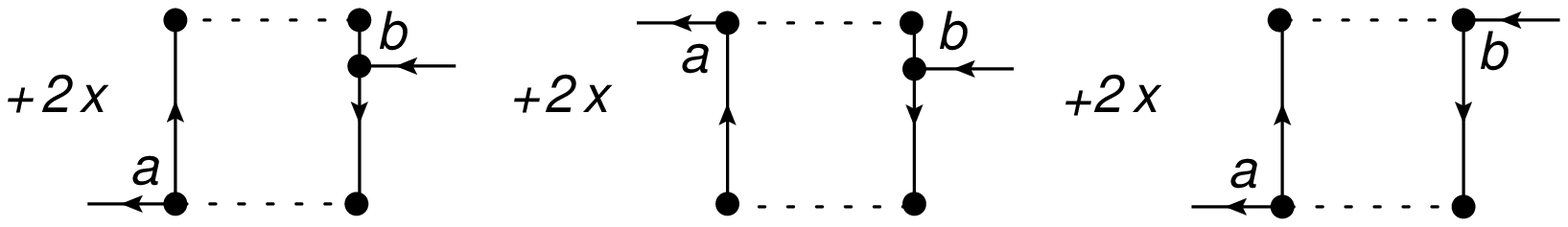}}
\\&
\raisebox{-15pt}{\includegraphics[width=220pt,height=40pt]{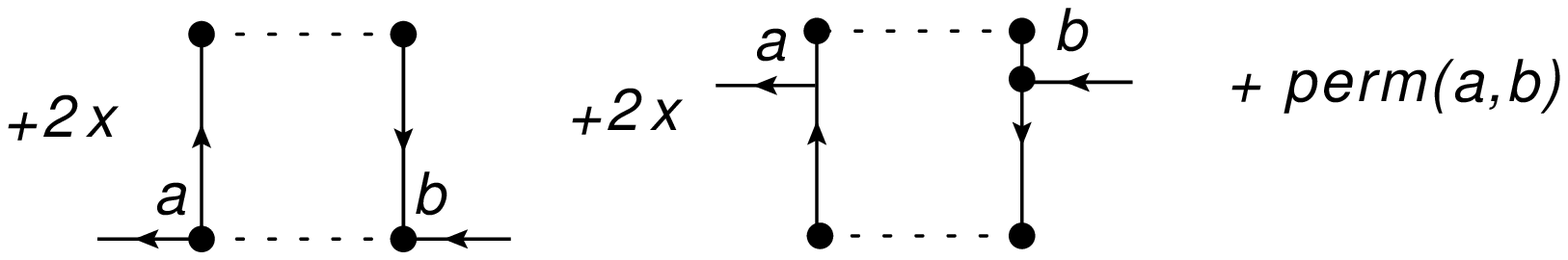}}
\Bigg),
\end{split}
\end{equation}
Explicitly written, the expression for $\beta_{\Delta}$ reads
\begin{equation}
\begin{split}
& \beta_{\Delta}=-\int_0^{\infty} \frac{dq}{\pi} 
\partial_t\widehat{R}_k(q)\Bigg\{-\widehat{G}_k(q;\phi_a)^3
\Big [\Delta_k(\phi_a, \phi_a) \times \\&
\Delta_k^{(2,0)}(\phi_a, \phi_b) + \Delta_k^{(1,0)}(\phi_a, 
\phi_b)\Big(\Delta_k^{(0,1)}(\phi_a, \phi_a) + \Delta_k^{(1,0)}(\phi_a, \phi_a) 
\\&
- 3\big [U^{(3)}_k(\phi_a) + q^{1 + 2\sigma} Y'_k(\phi_a)\big ]\Delta_k(\phi_a, 
\phi_a)\widehat{G}_k(q;\phi_a) \Big) \Big ]\\ &
- \Big [\Delta_k(\phi_a, \phi_b)\Delta_k^{(1,1)}(\phi_a, \phi_b) + 
\Delta_k^{(0,1)}(\phi_a, \phi_b)\Big(\Delta_k^{(1,0)}(\phi_a, \phi_b) 
\\&
- 2\big [U^{(3)}_k(\phi_a) + q^{1 + 2\sigma} Y'_k(\phi_a)\big ] 
\Delta_k(\phi_a, \phi_b) \widehat{G}_k(q;\phi_a) \Big) \Big ] \times 
\\& \widehat{G}_k(q;\phi_a)^2 \widehat{G}_k(q;\phi_b) 
+ \big [U^{(3)}_k(\phi_a) + q^{1 + 2\sigma} Y'_k(\phi_a)\big ]\Delta_k(\phi_a, 
\phi_b)\\&
\times  \Big(\Delta_k^{(0,1)}(\phi_a, \phi_b) - \big [U^{(3)}_k(\phi_b) + q^{1 
+ 2\sigma} Y'_k(\phi_b)\big ]
 \times \\&
\Delta_k(\phi_a, \phi_b) \widehat{G}_k(q;\phi_a) 
\Big)\widehat{G}_k(q;\phi_a)^2\widehat{G}_k(q;\phi_b)^2  + perm(a,b)\Bigg\} \,.
\end{split}
\end{equation}

\section{Toy model integral}
\label{toy_p}

We consider the following one-dimensional integral
\begin{equation}
 \label{toy1}
\mathcal I(p)= 
\int^{\infty}_{-\infty}dq\frac{1}{[|q+p|^\sigma+f((q+p)^2)][|q|^\sigma+f(q^2)]}
\end{equation}
with $1/3<\sigma<1/2$, in the vicinity of $p=0$. (As $\mathcal I(p)=\mathcal 
I(-p)$, we restrict ourselves to $p \geq 0$.) 
The function $f(q^2)$ is chosen such that $f(0)\neq 0$ and that its large 
$q^2$ behavior garantees the convergence of the integral. As a result $\mathcal 
I(p=0)$ is finite and we look 
for the leading $p$-dependence when $p\rightarrow 0$. 
A naive expansion in $p$ predicts a $p^2$ dependence, but this turns out to be 
wrong. To obtain the correct 
behavior, we first rewrite $\mathcal I(p)-\mathcal I(0)$ as
\begin{equation}
\begin{split}
\label{toy4}
&\mathcal I(p)-\mathcal I(0)=
\int^{\infty}_{-\infty}dq 
\\&\Bigg[\frac{1}{(|q+\frac{p}{2}|^\sigma+f((q+\frac{p}{2})^2))(|q-\frac{p}{2}
|^\sigma+f((q-\frac{p}{2})^2))}-
\\& 
\frac{1}{2}\Big(\frac{1}{[|q+\frac{p}{2}|^\sigma+f((q+\frac{p}{2})^2)]^2}+\frac{
1}{[|q-\frac{p}{2}|^\sigma+f((q-\frac{p}{2})^2)]^2}\Big)\Bigg]
\end{split}
\end{equation}
where, for convenience, we have expressed $\mathcal I(p=0)$ in a symmetrized 
form by changing the integration variable to 
$q\pm\frac{p}{2}$.

We now introduce the variable $x=\sqrt{q^2+(p^2/4)}$, so that
\begin{equation}
\begin{split}
 \label{denom_exp}
&|q\pm\frac{p}{2}|^\sigma+f(|q\pm\frac{p}{2}|^2)=f(x^2)+|x|^\sigma \pm 
\frac{qp}{|x|^{2-\sigma}}(\frac{\sigma}{2}+\\&
|x|^{2-\sigma}f'(x^2))
+\frac{q^2p^2}{2|x|^{4-\sigma}}(\frac{\sigma(\sigma-2)}{4}+|x|^{4-\sigma}
f''(x^2))+\cdots
\end{split}
\end{equation}
where the ellipses denotes terms with higher-order powers in $p$ (at fixed $x$).

After inserting Eq.~(\ref{denom_exp}) in Eq.~(\ref{toy4}) and changing variable 
from $q$ to $z=q/p$, we obtain
\begin{equation}
\begin{split}
\label{toy5}
&\mathcal I(p)-\mathcal I(0)=\\&
-\frac{p^{1+2\sigma}}{2}\int^{\infty}_{-\infty}dz\frac{z^2}{(z^2+\frac{1}{4})^{
2-\sigma}}\Big [\frac{\sigma^2
+{\rm O}(p^{2-\sigma})}{f(0)^4+{\rm O}(p^{\sigma})}\,,
\Big]
\end{split}
\end{equation}
so that
\begin{equation}
 \label{lowestp_I}
\lim_{p\rightarrow 0} \frac{\mathcal I(p)-\mathcal 
I(0)}{p^{1+2\sigma}}=-\frac{\sigma^2}{2f(0)^4}\int^{\infty}_{-\infty}dz\frac{z^2
}{(z^2+\frac{1}{4})^{2-\sigma}}\,,
\end{equation}
where the integral over $z$ converges in the range of $\sigma$ considered.

\section{Derivation of $\beta_{Y,an}$}
\label{app:betaY_an}

\begin{figure}[h!]
\begin{center}
\includegraphics[width=240pt,height=50pt]{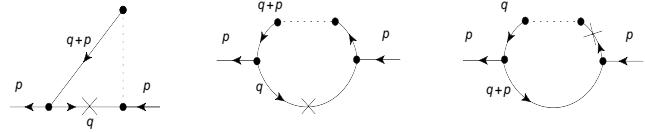}    
\caption{The diagrams contributing to the anomalous kinetic term in $\vert 
p\vert^{1+2\sigma}$. 
A cross denotes an insertion of $\partial_t\widehat R_k(q)$.}\label{fig2}
\end{center}
\end{figure}

Consider the diagrams shown in Fig. \ref{fig2}. They enter in the right-hand 
side of the 
flow equation of $\Gamma^{(2)}_{k1}(p;\phi)$ as
\begin{equation}
\begin{split}
\label{betY2start}
& \mathcal I_{an}( p)=-\frac{1}{2\pi}\int_{-\infty}^{+\infty} dq\, 
\partial_t\widehat R_k(q) 
\widehat{G}_k(q;\phi)^2\widehat{G}_k(p+q;\phi)
 \\&
 \times U'''_k(\phi)\Big (-2\big 
[\Delta^{(1,0)}_k(\phi,\phi)+\Delta^{(0,1)}_k(\phi,\phi) \big ]+ 
 \\&
 U'''_k(\phi) \Delta_k(\phi,\phi)\big 
[2\widehat{G}_k(q;\phi)+\widehat{G}_k(q+p;\phi)\big]\Big )
\end{split}
\end{equation}
with $\widehat{G}_k(q;\phi)$ given by Eq. (\ref{FT1exprssn}).

The expressions which need to be expanded in small $p$ (we consider $p\geq0$) 
are therefore the three integrals 
$\mathcal I_{an,1}( p)=\int dq\partial_t\widehat 
R_k(q)\widehat{G}_k(q;\phi)^2\widehat{G}_k(q+p;\phi)$, 
$\mathcal I_{an,2}( p)=\int dq\partial_t\widehat 
R_k(q)\widehat{G}_k(q;\phi)^3\widehat{G}_k(q+p;\phi)$ 
and $\mathcal I_{an,3}( p)=\int dq\partial_t\widehat 
R_k(q)\widehat{G}_k(q;\phi)^2
\widehat{G}_k(q+p;\phi)^2$. We follow exactly the same procedure as in the 
Appendix \ref{toy_p}: we symmetrize the integrands when needed, introduce the 
variable 
$x=\sqrt{q^2+\frac{p^2}{4}}$, expand in small $p$ and finally change the 
integration variable from $q$ 
to $z=q/p$. The outcome at leading order in $p$ is then  
\begin{equation}
\begin{split}
\label{stp}
& \mathcal I_{an,1}( p)- \mathcal I_{an,1}( 0)=\\& - 
p^{1+2\sigma}\partial_t\widehat R_k(0)\widehat{G}_k(0;\phi)^5
\int_{-\infty}^{+\infty}dz \frac{\sigma^2z^2}{(z^2+\frac{1}{4})^{2-\sigma}} \,,
\\&
 \mathcal I_{an,2}( p)- \mathcal I_{an,2}( 0)=\\&- 
p^{1+2\sigma}\partial_t\widehat R_k(0)\widehat{G}_k(0;\phi)^6
 \int_{-\infty}^{+\infty}dz\frac{3\sigma^2z^2}{2(z^2+\frac{1}{4})^{2-\sigma}} 
\,,
\\&
 \mathcal I_{an,3}( p)- \mathcal I_{an,3}( 0)=\\& - 
p^{1+2\sigma}\partial_t\widehat R_k(0)\widehat{G}_k(0;\phi)^6
\int_{-\infty}^{+\infty}dz \frac{2\sigma^2z^2}{(z^2+\frac{1}{4})^{2-\sigma}} \,.
\end{split}
\end{equation}
After inserting the above expressions in (\ref{betY2start}) and performing the 
integral over $z$, one immediately 
obtains Eq. (\ref{betY2}). 

\section{Recovering the perturbative result at order $\epsilon=\sigma-1/3$ from 
the 
NP-FRG}
\label{NPRGtoPERT}

We study the NP-FRG flow equations expressed in dimensionless quantities in the 
vicinity of the boundary value to classical (long-range) behavior, 
$\sigma_l=\frac{1}{3}$. The fixed point 
should be close to the Gaussian fixed point characterized by 
$u''_k=y_k=0$ and $\delta_k(\varphi,\delta\varphi)=1$ (recall that the 
long-range kinetic part is not renormalized so that $\eta=2-\sigma$). For 
$\sigma=1/3+\epsilon$, 
we expand the functions in powers of the fields,
\begin{equation}
\begin{split}
&u''_k(\varphi)=u_{k,2}+\frac 12 u_{k,4}\varphi^2 +{\rm O}(\varphi^4) \,,\\&
 y_k(\varphi)=\frac 12 y_{k,2}\varphi^2+{\rm O}(\varphi^4) \,,\\&
 \delta_k(\varphi,\delta\varphi)=\\& 1+\frac 12 
(\delta_{k,20}\varphi^2+\delta_{k,02}\delta\varphi^2) 
 +{\rm O}(\varphi^4,\delta\varphi^4,\varphi^2\delta\varphi^2)\,,
\end{split}
\end{equation}
where one expects $u_{k,2}$, $u_{k,4}$, $y_{k,2}$, $\delta_{k,20}$, 
$\delta_{k,02}$ to be at least of O($\epsilon$) 
around the fixed point.

After inserting the above expressions in the beta function 
$\widetilde{\beta}_{u''}(\varphi)$ obtained from 
Eq. (\ref{eq_beta_u''}), one finds the following flow equation for the coupling 
constant $u_{k,4}$: 
\begin{equation}
\label{flw_u4}
 \partial_t u_{k,4} = -3\epsilon u_{k,4} +36 u^2_{k,4} 
I_4(u_{k,2})+O(\epsilon^{2})
\end{equation}
where
\begin{equation}
\label{integral_I1}
I_n(u_2) = \frac{1}{2}
 \int_{0}^{\infty} \frac{dq}{\pi}\widetilde \partial_t s(q^2) \,  \widehat 
p_0(q)^n
\end{equation}
with $\widehat p_0(q)=[q^\sigma+ s(q^2) + u_{2}]^{-1}$ and $\widetilde 
\partial_t s(q^2)= 
\sigma s(q^2) - 2 q^2 s'(q^2)$. Eq. (\ref{flw_u4}) admits a nontrivial 
fixed-point solution 
$u^*_4=\epsilon/[12 I_4(0)]+O(\epsilon^{2})$. 

Similarly, the flow of $u_{k,2}$ reads
\begin{equation}
 \partial_t u_{k,2} =-\sigma u_{k,2} - 4 u_{k,4} I_3(u_{k,2})+{\rm 
O}(\epsilon^{2})\,,
\end{equation}
which leads to $u^*_2=- \epsilon I_3(0)/[3\sigma I_4(0)]+O(\epsilon^{2})$.

On the other hand, from the beta function for $\delta_k$ and the condition 
$\partial_t\delta_k(0,0)=0$, one derives that
\begin{equation}
2\eta-\overline{\eta}= O(\epsilon^{2})\,.
\end{equation}

To derive the $\epsilon$ dependence of the exponent $1/\nu$, one starts from 
the 
flow equation for $u_{k,2}$ and perturbs the fixed point by a small constant 
$k^{-1/\nu} \delta u_2$. By using 
that $I_3(u^*_2+\delta u_2)\approx I_3(u^*_2)-3 I_4(u^*_2)\delta u_2$ one finds 
the following 
linearized equation:
\begin{equation}
\frac{-1}{\nu} \delta u_2 =  (-\sigma +\epsilon)\delta u_2+O(\epsilon^{2}) \,,
\end{equation}
which implies, as $\sigma=1/3+\epsilon$,
\begin{equation}
 \label{nu_e1}
\frac{1}{\nu}=\frac 13+O(\epsilon^{2})\,.
\end{equation}
The above expression for $1/\nu$ is identical to the perturbative 
result.\cite{young_77}

Finally, we can also derive the expression for the eigenvalue $\lambda$. In the 
vicinity of 
the Gaussian fixed point we look for an eigenfunction $f\lambda(\varphi)$ of 
the form 
$f_\lambda(\varphi)=f_{\lambda,0} + (1/2)f_{\lambda,2} \varphi^2+\cdots$. We 
find that  
$f_{\lambda,2} ={\rm O}(\epsilon^2)$ and the eigenvalue equation then reads
\begin{equation}
\begin{split}
\lambda f_{\lambda,0}= (\frac 16-\epsilon) f_{\lambda,0} 
 + {\rm O}(\epsilon^2) \,,
\end{split}
\end{equation}
which leads to
\begin{equation}
\lambda=\frac 16 -\epsilon  + {\rm O}(\epsilon^2) \,.
\end{equation}
This is the result quoted in the main text.


\begin{thebibliography}{9}


\bibitem{imry-ma75}
Y. Imry and S. K. Ma, Phys. Rev. Lett. \textbf{35}, 1399 (1975).

\bibitem{nattermann98}
For 	a review, see T. Nattermann, \textit{Spin glasses and random fields} 
(World scientific, Singapore, 1998), p. 277.
 
 \bibitem{grin_76} 
 G.Grinstein, Phys. Rev. Lett. \textbf{37}, 944 (1976).
 
 \bibitem{young_77} 
 A.P.Young, J. Phys. C \textbf{10}, L257 (1977).
 
 \bibitem{parisi_79} 
 G.Parisi and N. Sourlas, Phys. Rev. Lett. \textbf{43}, 744 (1979).
 
 \bibitem{imbrie_84} 
 J.Z. Imbrie: Phys. Rev. Lett. 53, 1747.
  
 \bibitem{brich_kup_87} 
 J.Brichmont and A. Kupiainen, Phys. Rev. Lett. \textbf{59}, 1829 (1987).
 
 \bibitem{tarjus04} 
 G. Tarjus and M. Tissier, Phys. Rev. Lett \textbf{93}, 267008 (2004); Phys. 
Rev. B \textbf{78}, 024203 (2008).

 \bibitem{tissier06} 
 M. Tissier and G. Tarjus, Phys. Rev. Lett. {\bf 96}, 087202 (2006); 
\textbf{78}, 024204 (2008).
 
 \bibitem{tissier11} 
 M. Tissier and G. Tarjus, Phys. Rev. Lett. \textbf{107}, 041601 (2011); Phys. 
Rev. B \textbf{85}, 104202 (2012);  Phys. Rev. B \textbf{85}, 104203 (2012).
 
 \bibitem{fisherFRG}
D. S. Fisher, Phys. Rev. Lett. {\bf 56}, 1964 (1986).
O. Narayan and D. S. Fisher, Phys. Rev. B {\bf 46}, 11520 (1992).

\bibitem{FRGledoussal-chauve}
P. Le Doussal, K. J. Wiese, and P. Chauve, Phys. Rev. B \textbf{66}, 174201 
(2002); Phys. Rev. E \textbf{69}, 026112 (2004).

\bibitem{FRGledoussal-wiese}
P. Le Doussal and K. J. Wiese, Phys. Rev. E \textbf{79}, 051106 (2009); Phys. 
Rev. E \textbf{85}, 061102
(2011). 
  
 \bibitem{bacz_LR3d} 
 M. Baczyk, M. Tissier, G. Tarjus, and Y. Sakamoto, Phys. Rev. B \textbf{88}, 
014204 (2013).

 \bibitem{tarj_avlnch} 
 G. Tarjus, M. Baczyk, and M. Tissier, Phys. Rev. Lett. \textbf{110}, 135703 
(2013).
 
 \bibitem{dewenter} 
 T.Dewenter and A.K.Hartmann, arXiv:1307.3987.
 
 \bibitem{leuzzi} 
 L. Leuzzi and G. Parisi, Phys. Rev. B \textbf{88}, 224204 (2013).
 
 \bibitem{Bray} 
 A. J. Bray: J. Phys. C \textbf{19}, 6225 (1986).
 
 \bibitem{cassandro09} 
 M. Cassandro, E. Orlandi, and P. Picco, Commun. Math. Phys. \textbf{288}, 731 
(2009).
  
  \bibitem{rodgers-bray} 
 G. J. Rodgers and A. J. Bray: J. Phys. A: Math. Gen. \textbf{21}, 2177 (1988).

  \bibitem{monthus11} 
 C. Monthus and T. Garel, J. Stat. Mech., P07010 (2011).
 
 \bibitem{Par_book} M.Mezard, G.Parisi and M.A.Virasoro: \textit{Spin Glass 
Theory and Beyond}, World Scientific (1987).
 
 \bibitem{Ber02} 
 J.Berges, N.Tetradis and C.Wetterich, Phys. Rep. \textbf{363}, 223 (2002).

 \bibitem{wetterich93} 
 C. Wetterich, Physics Letters B \textbf{301}, 90 (1993).

\bibitem{litim_opt}
D. F. Litim, Phys. Lett. B \textbf{486}, 92 (2000); Nucl. Phys. B \textbf{631}, 
128 (2002). 	

\bibitem{canet_opt}
L. Canet, B. Delamotte, D. Mouhanna, and J. Vidal, Phys.
Rev. D \textbf{67}, 065004 (2003). 36	

\bibitem{pawlowski_opt}
J. M. Pawlowski, Ann. Phys. \textbf{322}, 2831 (2007).

 \bibitem{newt_rasp} 
 See for instance \textit{Numerical recipes}, Cambridge University press 
(1992), 
vol 1, sec. 9.4.
  
 \bibitem{balog_FP} 
 M. Baczyk, G. Tarjus, M. Tissier and I. Balog, J. Stat Mech., P06010 (2014).

 \bibitem{wu-machta05}
Y. Wu and J. Machta, Phys. Rev. Lett.  \textbf{95}, 137208 (2005);
Phys. Rev. B \textbf{74}, 064418 (2006).

 \bibitem{liu-dahmen07}
Y. Liu and K. A. Dahmen, Phys. Rev. E  \textbf{76}, 031106 (2007).

\bibitem{BBM96}
L. Balents L, J.-P. Bouchaud and M. Mezard, J. physique I \textbf{6}, 1007 
(1996).
 
 \bibitem{aval_FRG}
 A. A. Middleton, P. Le Doussal and K. J. Wiese, Phys. Rev. Lett. \textbf{98}, 
155701 (2007); 
P. Le Doussal, A. A. Middleton, and K. J. Wiese, Phys. Rev. E \textbf{79}, 
050101 (2009).
 
 \end{thebibliography}
\end{document}